\documentclass[prl,reprint,amsmath,amssymb,aps]{revtex4-1}
\pdfoutput=1
\usepackage{fullpage}
\usepackage{graphicx}
\usepackage{graphicx}
\usepackage{dcolumn}
\usepackage{bm}
\newcommand{\be}{\begin{equation}}
\newcommand{\ee}{\end{equation}}
\newcommand{\bea}{\begin{eqnarray}}
\newcommand{\eea}{\end{eqnarray}}
\newcommand{\bl}{\mbox{\boldmath $\ell$}}
\newcommand{\bnabla}{\mbox{\boldmath $\nabla$}}
\renewcommand{\epsilon}{\varepsilon}

\begin{document}
\preprint{\em DCPT-16/xx}
\title{
  Untangling knots via reaction-diffusion dynamics of vortex strings
}
\author{Fabian Maucher$^{\dagger\star}$ and Paul Sutcliffe$^\star$\\ \ }
\affiliation{
  $^\dagger$Joint Quantum Centre (JQC) Durham-Newcastle, Department of Physics,
Durham University, Durham DH1 3LE, United Kingdom.\\
  $^\star$Department of Mathematical Sciences,
Durham University, Durham DH1 3LE, United Kingdom.\\ 
Email: fabian.maucher@durham.ac.uk, \ p.m.sutcliffe@durham.ac.uk}
\date{April 2016}

\begin{abstract}
  We introduce and illustrate a new approach to the unknotting problem via the dynamics of vortex strings in a nonlinear partial differential equation of reaction-diffusion type. To untangle a given knot, a Biot-Savart construction is used to initialize the knot as a vortex string in the FitzHugh-Nagumo equation.
Remarkably, we find that the subsequent evolution preserves the topology of the knot and can untangle an unknot into a circle. Illustrative test case examples are presented, including the untangling of a hard unknot known as the culprit.
Our approach to the unknotting problem has two novel features, in that it
applies field theory rather than particle mechanics
and uses reaction-diffusion dynamics in place of energy minimization.
\end{abstract}
\maketitle
The fundamental problem in knot theory is to determine whether a closed loop embedded in three-dimensional space is knotted. This unknotting problem, and more generally the identification of knots, leads to deep mathematical connections throughout the natural sciences from quantum field theory \cite{Wi} to the properties of DNA \cite{WC}.
There has been considerable recent interest in knotted vortex strings
in field theory in a range of areas from fluid dynamics \cite{KI}
to ultracold atoms \cite{hall}, with superfluid vortex strings
providing a canonical example. However, in such systems there are reconnection events (where crossing strings exchange partners) which change the knot type and eventually all knots completely untie, accompanied by the production of a set of unknotted circles.
A very recent comprehensive analysis of the untying pathways of initially knotted superfluid vortex strings suggests that some universal mechanisms are at work \cite{KKI}.
An exception to this class of field theories is the Skyrme-Faddeev model \cite{FN}, where static minimal energy knotted vortex strings exist \cite{BS}, stabilized by a topological charge. However, even in this system, reconnection events take place, therefore the knot type is not conserved during energy relaxation, and indeed an optimal minimal energy knot can appear via reconnection from the unknot.
\begin{figure}[ht]\begin{center}\includegraphics[width=6.5cm]{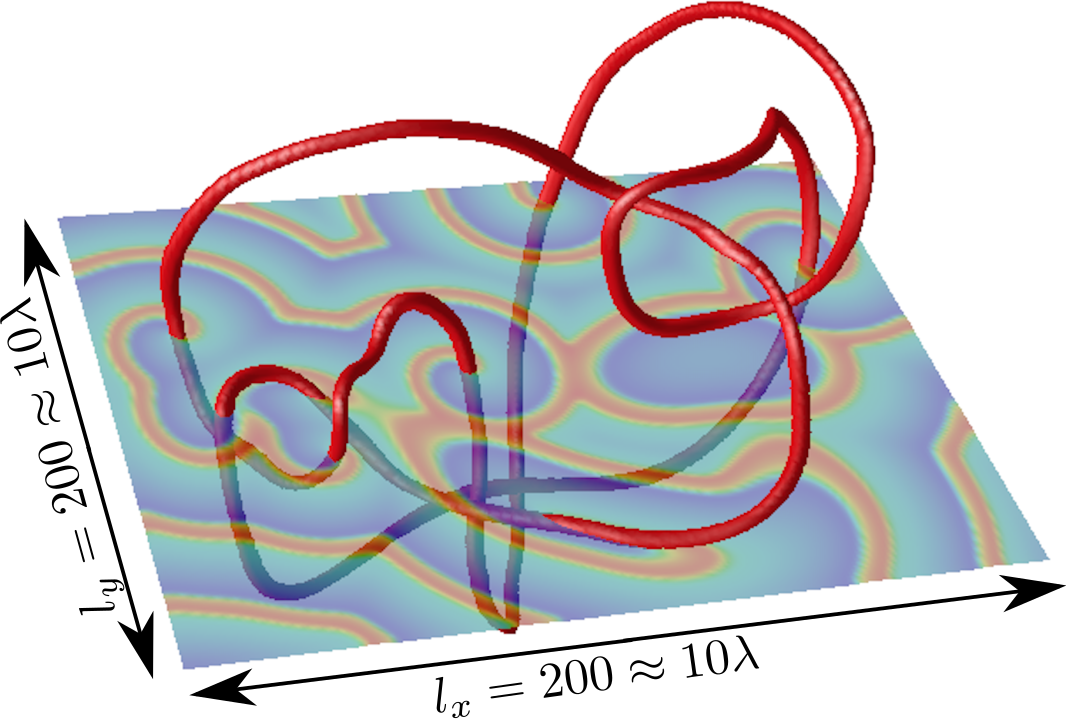}    \caption{The vortex string of the culprit unknot and the field $u$ in a planar slice.}\label{fig-culpritx}  \end{center}\end{figure}

The ubiquitous reconnection of vortex strings in nonlinear physical fields has
reduced their impact on knot theory. In the present paper we propose
a novel application of vortex strings to knots by considering a
reaction-diffusion system in which the vortex strings appear to be averse to reconnection. In particular, we present some illustrative test case examples of unknots that untangle without reconnection, to yield a circular unknot.
The dynamics of vortex strings in this system is therefore in remarkable
contradistinction to the universal properties discussed in \cite{KKI}.
Studying knots via topology preserving vortex string dynamics is a new
paradigm connecting classical field theory, knot theory and partial differential equations. In contrast to conventional knot untangling methods, which apply particle mechanics and energy minimization,  our approach is novel in its application of field theory and reaction-diffusion dynamics to this problem.
The use of reaction-diffusion systems as a computation device is a current hot
topic in the field of novel and emerging computing paradigms \cite{ADA} and
our approach to knot untangling provides a new application in this field.
\begin{figure*}\begin{center}\includegraphics[width=16cm]{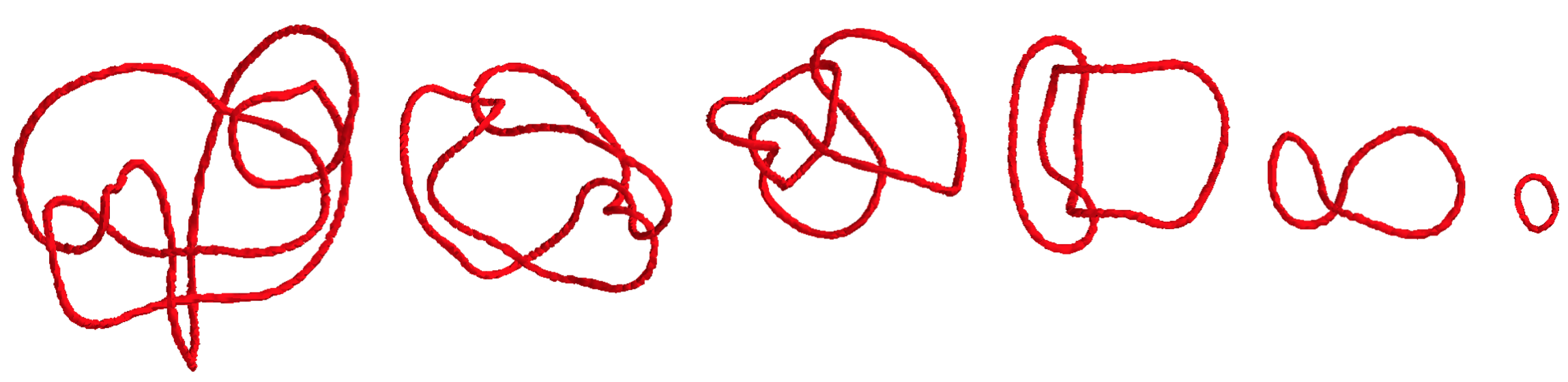}\caption{The vortex string at $t=60, 1000, 2000, 3000, 4000, 4600$, untangling the culprit unknot.}\label{fig-culprit}\end{center}\end{figure*}

The reaction-diffusion equation of interest in this paper is the FitzHugh-Nagumo system, which is a simple mathematical model of cardiac tissue as an excitable medium \cite{Winfree}. The nonlinear partial differential equations are given by
\be
\frac{\partial u}{\partial t}=\frac{1}{\epsilon}(u-\frac{1}{3}u^3-v)+\nabla^2 u,
\quad
\frac{\partial v}{\partial t}=\epsilon(u+\beta-\gamma v),
\ee
where $u({\bf r},t)$ and $v({\bf r},t)$ are the real-valued physical fields defined throughout three-dimensional space with coordinate ${\bf r}$,
and where $t$ denotes time.
The remaining variables are constant parameters that we fix to be
$\epsilon=0.3, \ \beta=0.7, \ \gamma=0.5$ from now on.

In two-dimensional space this system, with the parameter values given above,
has rotating vortex solutions, often called spiral waves \cite{Winfree},
with a period $T=11.2$ and $u$ and $v$ wavefronts in the form of an involute spiral with a wavelength $\lambda=21.3.$ Characteristic time and length scales are determined by the parameters $T$ and $\lambda$ and in particular the vortex core can be assigned a radius $\frac{\lambda}{2\pi}$.
The centre of the vortex
is the point at which $|{\bnabla} u\times \bnabla v|$ is maximal,
and this quantity is 
localized in the vortex core.

Motivated by the observed short-range repulsive force between vortex cores, it was conjectured many years ago \cite{StWi} that knotted vortex strings in the three-dimensional system might preserve their topology under FitzHugh-Nagumo evolution. Numerical support for this conjecture was obtained \cite{SW} by simulations on long time scales (thousands of vortex rotation periods) for the simplest knot and link (the trefoil knot and the Hopf link). However, investigations of more complicated vortex string geometries or potential applications to knot theory have remained elusive because of the difficulties in obtaining more general initial conditions beyond symmetric representations of simple knots.

In this paper, we remove this obstacle by adapting a recent scheme \cite{Sa}, introduced for superfluid vortices, to the FitzHugh-Nagumo system. The method involves a Biot-Savart construction and allows the creation of initial conditions for a vortex string
with arbitrary shape and topology.
In detail, given any non-intersecting closed curve $K$, one imagines this curve to be a wire carrying a constant current and computes the associated magnetic field ${\bf B}({\bf r})$
using the Biot-Savart law
\be
   {\bf B}({\bf r})=\frac{1}{2}\int_{K}\frac{({\bf r}-\bl)\times d\bl}{|{\bf r}-\bl|^3},
   \ee
   where $\bl$ is a coordinate on $K$. The scalar potential $\Phi({\bf r})$, defined by ${\bf B}=\bnabla {\Phi}$ is then computed by fixing a base point ${\bf r}_0$ and performing the line integral
   \be
   \Phi({\bf r})=\int_C {\bf B}({\bf r}')\cdot d{\bf r}',
   \ee
   where $C$ is a curve 
   that starts at ${\bf r}_0$ and ends at ${\bf r}$ whilst avoiding $K.$ Finally, the initial FitzHugh-Nagumo fields $u$ and $v$ are obtained from the scalar potential using the formulae
   \be
   u=2\cos\Phi-0.4, \quad v=\sin\Phi-0.4.
   \label{ic}
   \ee
   The subtraction of the constants in (\ref{ic}) is motivated by the fact that the centre of a vortex is roughly associated with the field values $(u,v)=(-0.4,-0.4)$. The scaling factor of 2 reflects the property that the fields outside the core of a single planar vortex perform a periodic oscillation in the $(u,v)$-plane with a range in the $u$ direction that is roughly twice that in the $v$ direction.  

\begin{figure}
  \begin{center}
    \includegraphics[width=8cm]{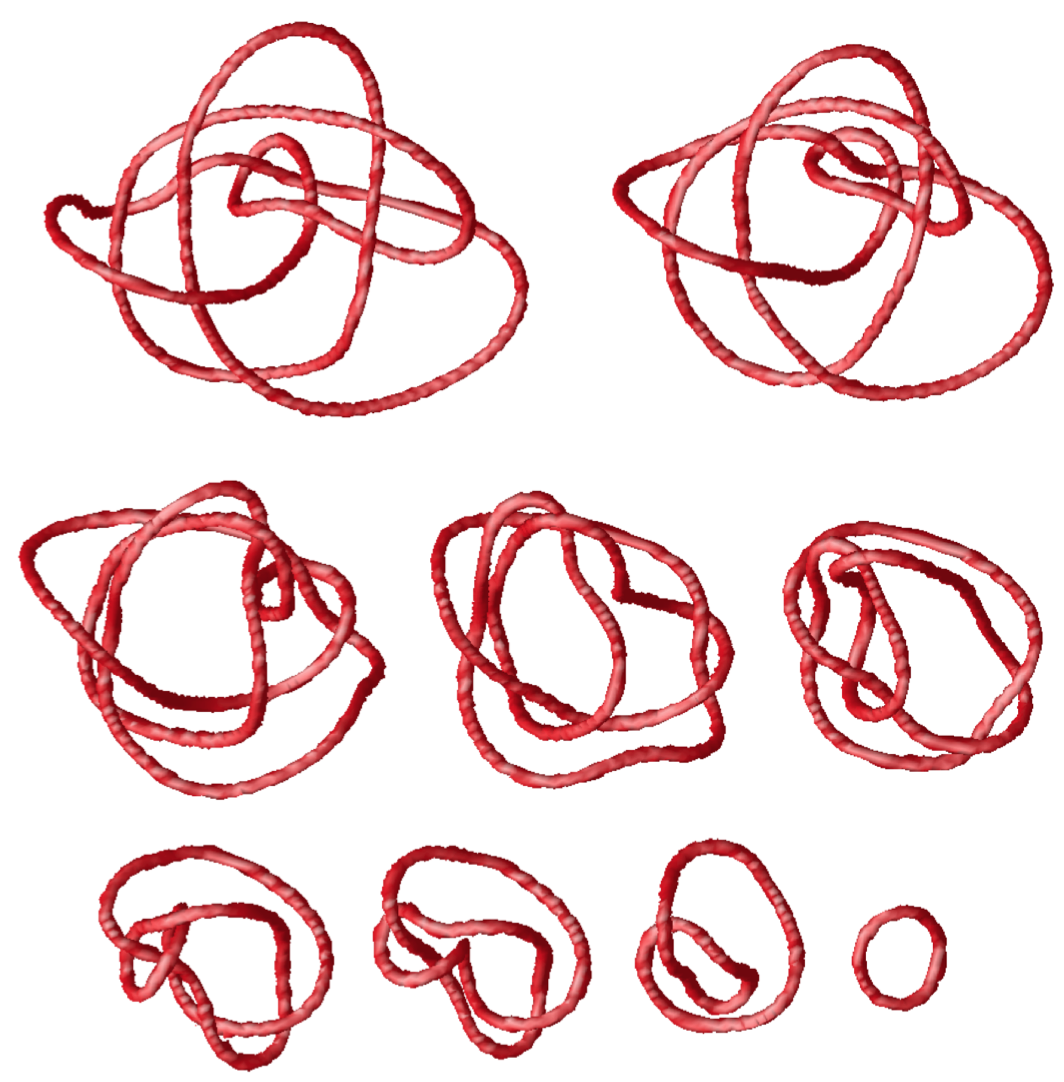}
    \caption{A vortex string is displayed at the times $t=200, 500, 900, 1200, 1500, 1800, 1900, 2100, 2500$, during the untangling of a 13-crossing unknot.}
\label{fig-o13}
  \end{center}
\end{figure}

After a time scale of the order of a vortex rotation period $T$,
the above construction generates a vortex string along $K$.
Fig.~\ref{fig-culpritx} displays an example numerical
implementation of this construction in a region of size
$200\times 200\times 150$ and after a time $t=60$, where the
FitzHugh-Nagumo equations are evolved numerically using standard methods with
 Neumann boundary conditions.
 The vortex string is displayed by plotting the isosurface where $|{\bnabla} u\times \bnabla v|=0.1$, and the same method is used to display all the vortex strings presented in this paper.  The field $u$ in a planar slice is also displayed,
 to illustrate the spiral wavefronts that emanate from the  vortex string.

 Fig.~\ref{fig-culprit} displays the knot already shown in Fig.~\ref{fig-culpritx}, but more clearly as a projection without the $u$ field in the
$(x,y)$-plane. This knot has 10 crossings in the given projection but is in fact an unknot known as the culprit \cite{KL}. It is an example of a hard unknot, which means that the number of crossings must first be increased in the process of untangling this unknot into a circle.
 The resulting evolution is shown (to scale) in the remainder of Fig.~\ref{fig-culprit} and is available as the movie
 file {\em culprit.mpg} in the supplementary material. This result demonstrates that FitzHugh-Nagumo dynamics successfully untangles this unknot, preserving the topology whilst evolving the geometry to the circular unknot. It is the first example, in any field theory, of vortex string dynamics that untangles a knot and is free from reconnection.
This field theory method seems at least as powerful 
as the other three-dimensional untangling strategies, such as energy minimization, all of 
which seem significantly more powerful than combinatorial untangling 
strategies, which are known to have trouble with configurations like 
the culprit. 
 
 A second unknot example, with more initial crossings than the culprit, is displayed (to scale) in Fig.~\ref{fig-o13}, using the same size grid. This figure
 illustrates the process in which this unknot with 13 initial crossings is successfully untangled without reconnection. The evolution is available as the movie {\em unknot13.mpg} in the supplementary material.

\begin{figure*}[ht]
  \begin{center}
    \includegraphics[width=13.0cm]{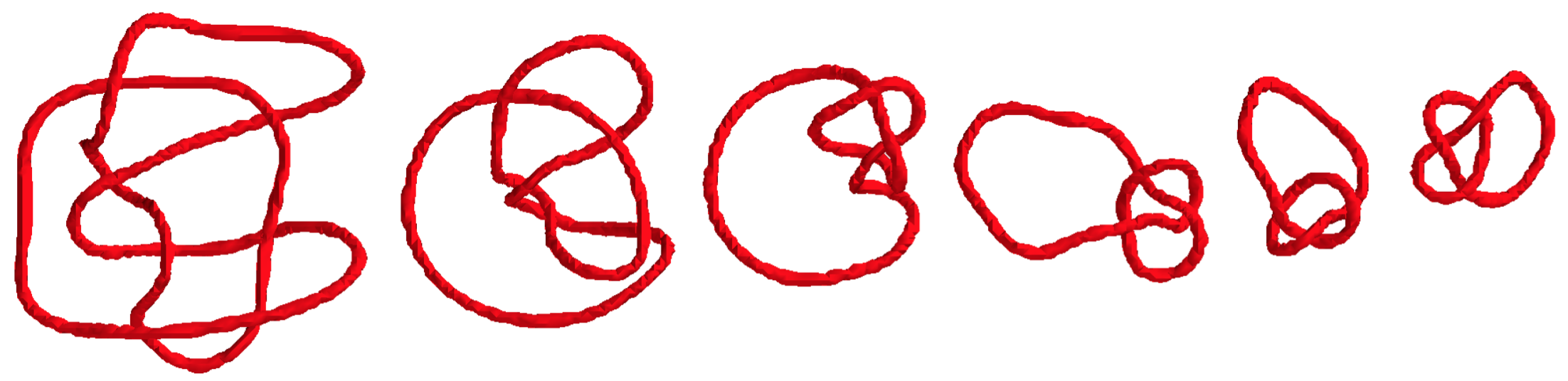}
    \caption{The vortex string at $t=60, 400, 760, 1600, 2200, 2720$, illustrating the simplification of a trefoil knot.}
\label{fig-cravat}
  \end{center}
\end{figure*}

As a final test of our new approach, we consider
the knot displayed in the first image in Fig.~\ref{fig-cravat}.
The initial knot has 7 crossings in this projection and, as every gentleman will recognize, this is a knot used to tie a cravat, if the two loose ends are then joined. The subsequent evolution is displayed (to scale) in the remaining images in Fig.~\ref{fig-cravat}
and is available as the movie {\em cravat.mpg} in the supplementary material.
Again there are no reconnections and the dynamics simplifies the geometry of the knot to reveal a trefoil knot with 3 crossings.

Vortex string reconnections are certainly not forbidden in the FitzHugh-Nagumo model and we have been able to force reconnections by initializing a vortex string with segments that are initially separated by less than a core diameter, so that reconnection takes place before the rotating spiral wave vortices have completely formed.
However, the dynamics appears to be averse to creating this situation from fully formed vortices that have attained separations beyond a vortex core diameter (as illustrated in Fig.~\ref{fig-culpritx}). By a suitable choice of scale, any given knot can be initialized so that all segments are separated by more than a core diameter, as we have done for the illustrative test case examples presented in this paper.  
For these, and a number of similar examples that we have investigated for 
a variety of initial knots, we find no evidence of string reconnection, providing the initial knot has a minimal distance between segments that is greater than  a vortex core diameter.

The examples we have studied so far are simple enough that the absence of string collisions and reconnections can be verified manually, but to investigate the untangling of more complex examples,
such as the notoriously difficult Ochiai unknot \cite{Oc}, would
require the implementation of some form of automatic string collision detection, to be certain that the knot topology is preserved. It could be that sufficiently complicated knots, beyond those studied so far, are not immune to vortex string reconnection, and it would be interesting to investigate this further and determine which, if any, types of knot are capable of inducing this phenomenon.

Methods have been developed to reduce the field theory dynamics to an effective
dynamics of the vortex string \cite{Ke,BHZ}, but they apply only in the regime of slight curvature and twist. It is therefore a challenging open problem to develop an effective string dynamics that is capable of reproducing the untangling motion described in the present paper and to explain the aversion to string reconnection.  

Simulated annealing methods have been developed in mechanical models of knots
\cite{HGK} to resolve the problem of the knot conformation becoming trapped
in a local energy minimum.
As our approach is based on reaction-diffusion evolution, rather than energy minimization, there is no issue regarding the knot conformation becoming trapped in a local energy minimum. The relevant question for reaction-diffusion evolution
is whether the flow has multiple attractors that can trap the knot conformation or if there is only one unknotted attracting steady state. This is an interesting mathematical problem that requires further study to determine whether this approach is an improvement on energy minimization methods, with regard to trapped conformations.
As untangling via reaction-diffusion evolution appears to explore very different untangling pathways and knot geometries to energy minimization, this may lead to some new understanding of the properties of physical knots, such as those found in DNA, which may also utilize chemical or other messengers to exchange information between different arcs of the knot, as modelled by wave interactions in reaction-diffusion dynamics.

By adapting a new construction of vortex string initial conditions to the FitzHugh-Nagumo equation, we have been able to investigate the
reaction-diffusion evolution of complex vortex string geometries with arbitrary topology. Remarkably, in contrast to other field theories, we find that
the vortex strings
preserve their knot topology and evolve without reconnection
to untangle a knot into a simplified geometrical form.
The fact that very complicated initial geometries untangle without reconnection is unexpected and generates a new open problem:
to explain how reaction-diffusion dynamics, which has no conserved quantities,
is able to preserve the knot topology whilst simultaneously simplifying the
geometry.
The implications of an understanding of this behaviour could have relevance in a  range of topics in physics, chemistry, biology and mathematics.

The theory of gradient flows for energy functionals on knots
has various significant technical challenges even to simply construct equations for the flow and to prove that short time solutions exist (see for
example the work \cite{He} on the gradient flow for the M\"obius energy).
A significant advantage of the reaction-diffusion flow of vortex strings is the absence of all these difficulties.

Finally, there are several chemical and biological systems
supporting spiral wave vortices,
from the oscillating Belousov-Zhabotinsky redox reaction to the chemotaxis of slime mould. All are described by similar reaction-diffusion equations that
possess vortex strings, so it is not beyond the realms of possibility that
in the future a chemical or biological system might be engineered that could 
untangle knots.\\

  {\bf \small \qquad \qquad ACKNOWLEDGEMENTS}
  
\noindent  We thank Hayder Salman and Lauren Scanlon for useful discussions.
This work is funded by the
Leverhulme Trust Research Programme Grant RP2013-K-009, SPOCK: Scientific Properties Of Complex Knots, and the STFC grant ST/J000426/1.

\end{document}